\documentclass[5p]{elsarticle}
\biboptions{sort&compress}

\usepackage{graphicx,amsmath,amssymb,slashed}

\journal{Physics Letters B}

\newcommand{\tttt}{t\bar t t \bar t}
\newcommand{\ourmS}{m_S}
\newcommand{\ourat}{a_t}
\def\be{\begin{equation}}
\def\ee{\end{equation}}
\def\bsp#1\esp{\begin{split}#1\end{split}} 
\newcommand{\amc}{{\sc MadGraph5\textunderscore}a{\sc MC@NLO}}

\begin{document} 

\begin{frontmatter}

\title{Probing top-philic sgluons with LHC Run I data}

\author[bristol,vub,iihe]{Lana Beck}
\ead{lana.beck@cern.ch}
\author[vub,iihe]{Freya Blekman}
\ead{freya.blekman@vub.ac.be}
\author[iihe,ulb,gent]{Didar Dobur}
\ead{didar.dobur@cern.ch}
\author[strasbourg]{Benjamin Fuks}
\ead{benjamin.fuks@iphc.cnrs.fr}
\author[vub,iihe]{James Keaveney}
\ead{james.keaveney@vub.ac.be}
\author[vub,iihe,solvay]{Kentarou Mawatari}
\ead{kentarou.mawatari@vub.ac.be}

\address[bristol]{University of Bristol, HH Wills Physics Laboratory
 Tyndall Avenue, BS8 1TL United Kingdom}
\address[vub]{Vrije Universiteit Brussel, Pleinlaan 2, B-1050 Brussels, Belgium}
\address[iihe]{Interuniversity Institute for High Energies, Pleinlaan 2, B-1050 Brussels, Belgium}
\address[ulb]{Universit\'e Libre de Bruxelles, Campus de la Plaine, boulevard du Triomphe, B-1050 Brussels, Belgium}
\address[gent]{Universiteit Gent, Proeftuinstraat 86, B-9000 Gent, Belgium}
\address[strasbourg]{Institut Pluridisciplinaire Hubert Curien/D\'epartement
    Recherches Subatomiques, Universit\'e de Strasbourg/CNRS-IN2P3,\\
    23 rue du Loess, F-67037 Strasbourg, France}
\address[solvay]{International Solvay Institutes, Pleinlaan 2, B-1050
 Brussels, Belgium}

\begin{abstract}
Many theories beyond the Standard Model predict the existence of colored scalar states, known as sgluons, lying in the adjoint representation of the QCD gauge group.
In scenarios where they are top-philic, sgluons
are expected to be copiously pair-produced at the LHC via strong
interactions with
decays into pairs of top quarks or gluons. 
Consequently, sgluons can be sought in
multijet and multitop events at the LHC. We revisit two LHC Run I analyses
in which events featuring either the same-sign dileptonic
decay of a four-top-quark system or its single leptonic decay
are probed. Adopting a simplified model approach,
we show how this reinterpretation allows us to extract simultaneous bounds on the
sgluon mass and couplings.
\end{abstract}


\end{frontmatter}

\section{Introduction}
\label{sec:intro}

Despite its success in describing all experimental high-energy physics data, the
Standard Model (SM) of particle physics leaves many important and conceptual issues
unanswered. As a consequence, many theoretical frameworks extending
it have been developed over the last decades, and new phenomena have been
searched for experimentally. Weak scale supersymmetry, and in
particular its minimal realization known as the Minimal Supersymmetric Standard
Model (MSSM)~\cite{Nilles:1983ge,Haber:1984rc}, is one of the most studied of those
beyond the SM setups. It is however more and more constrained by
data, and especially by the recent results of the LHC
experiments~\cite{atlassusy,cmssusy}. There are nevertheless large varieties of
alternative non-minimal supersymmetric models that deserve to be investigated
and whose signatures may be different from the expectations of the minimal
choice.

Along these lines, we focus on $N\!=\!1\!/\!N\!=\!2$ hybrid~\cite{Fayet:1975yi,%
AlvarezGaume:1996mv} and $R$-symmetric~\cite{Salam:1974xa,Fayet:1974pd,%
Kribs:2007ac} supersymmetric theories that
both predict extra scalar partners to the SM gauge bosons. These
additional degrees of freedom lie in the adjoint representation of the gauge
group and are indeed not present in the MSSM.
Among the new particles, the colored states commonly dubbed sgluons have
received special attention as they are expected to be copiously produced at
hadron colliders~\cite{Plehn:2008ae,Choi:2008ub,Choi:2009jc,Schumann:2011ji,%
Kotlarski:2011zz,Kotlarski:2011zza,GoncalvesNetto:2012nt,Calvet:2012rk,Degrande:2014sta}.
Those fields however appear not only in
supersymmetry but also in vector-like confining
theories~\cite{Kilic:2009mi} and extra-dimensional models~\cite{Burdman:2006gy}.
Motivated by typical sgluon signatures that are similar in all these models, we
adopt a simplified model approach describing the dynamics of a scalar field
lying in the octet representation of $SU(3)_c$ and interacting with the
SM~\cite{Brooijmans:2012yi,Calvet:2012rk}. This subsequently
allows both for a model-independent approach and a simplification of the
non-minimal supersymmetric parameter spaces that in general contain hundreds of
free parameters.

This simplified approach has been used experimentally in order to search for hints of 
sgluons in LHC collision data at center-of-mass energies of 7 TeV and 8 TeV.
As a consequence of null results, limits on
the sgluon mass, $\ourmS$, have been extracted after probing
the production of a sgluon pair that decays into either a four top-quark system
yielding a same-sign dilepton final state~\cite{TheATLAScollaboration:2013jha},
or a four-jet final state~\cite{ATLAS:2012ds}. In the former case, $\ourmS$ is
bound to be larger than 800~GeV when we assume that the sgluon always decays
into
a top-antitop system. In the latter case, limits turn out to be weaker ($\ourmS
\gtrsim 300$~GeV) and are obtained after assuming that the sgluon always decays
into a dijet state. Additionally, stronger constraints could be derived in the
context of single sgluon production, the sgluon mass being pushed in this case
above the multi-TeV scale~\cite{Aad:2011fq,Chatrchyan:2013qha}. These bounds are
however very model-dependent and could be evaded as soon as the sgluon is
allowed to couple to the top quark, as in realistic ultraviolet-complete setups
such as those mentioned. We therefore consider a framework where
sgluons can couple to both quarks and gluons, and then revisit
constraints on sgluon simplified models by reinterpreting recent LHC analyses of
all data recorded at a collision center-of-mass energy of
8~TeV. More precisely, we consider two CMS studies of four-top-quark topologies,
a first one focusing
on same-sign dilepton events~\cite{Chatrchyan:2013fea} and a second one on
single lepton events~\cite{Khachatryan:2014sca}. We hence derive, for
the first time, limits on the sgluon mass and 
coupling strengths to the SM particles simultaneously.

The rest of this paper is organized as follows. In Section~\ref{sec:model}, we
briefly describe our simplified theoretical framework for sgluon phenomenology at the
LHC, and present the sgluon mass dependence of both the total sgluon-pair
production rate and the sgluon branching ratios.
The reinterpretation of the LHC analyses of
Refs.~\cite{Khachatryan:2014sca,Chatrchyan:2013fea} is detailed in
Section~\ref{sec:lhc}, and our conclusions are presented in
Section~\ref{sec:conclusion}.

\section{A simplified model for top-philic sgluon phenomenology}
\label{sec:model}

In our study of sgluon production and decay at the LHC, we rely on a minimal
extension of the Standard Model allowing for a general description of sgluon
dynamics. To this aim, we construct a simplified model in which
we supplement the Standard Model by a single real color-octet scalar field
$S^a$, the superscript $a$ indicating an adjoint color index. The
kinetic and mass terms associated with this field
are given by the Lagrangian
\be
 {\cal L}= \frac{1}{2}D_{\mu}S^aD^{\mu} S_a
          -\frac{1}{2}\ourmS^2 S^a S_a\ ,
\ee
which includes the gauge interactions of a sgluon pair to gluons through the
QCD-covariant derivative,
\be
 D_{\mu}S^a= \partial_{\mu} S^a + g_s f_{bc}{}^a G_\mu^b S^c \ .
\ee
In our conventions, $g_s$ denotes the strong coupling
constant, $f_{bc}{}^a$ the structure
constants of the $SU(3)_c$ gauge group and $G_\mu^a$ represents
the gluon field.

Furthermore, in order to allow the sgluon for singly coupling to the Standard
Model degrees of freedom, we introduce the effective Lagrangian
\be
 {\cal L}_{\rm eff}= \bar t\, T_a (\ourat^L P_L + \ourat^R P_R)\, t\, S^a
 +\frac{a_g}{\Lambda} d_a{}^{bc} S^a F_{\mu\nu, b} F^{\mu\nu, c}
 + {\rm h.c.} 
\label{eq:SingleSgluLag}\ee
Its first term consists of dimension-four interactions of the sgluon with a pair
of top-antitop quarks whose left-handed and right-handed
coupling strengths are denoted by $\ourat^L$ and $\ourat^R$, respectively.
The second term of ${\cal L}_{\rm eff}$
models single sgluon interactions to gluons through a dimension-five operator,
the dimensionless coupling strength $a_g$ being
suppressed by the theory cutoff energy scale
$\Lambda$. In our notations, the $T_a$ matrices stand for the
generators of $SU(3)_c$ in the fundamental representation, while $d_a{}^{bc}$
are the symmetric structure constants of the group and $P_{L/R}$ denote the
chirality projection operators. Although sgluons can in
principle also couple in a flavor-changing-neutral way to different quark
species, we
only retain their flavor-conserving interactions with top quarks.
While not general, this choice is motivated by minimally flavor-violating
$R$-symmetric supersymmetric models where single sgluon interactions are
loop-induced by squarks and gluino in a way such that
only interactions
involving a pair of gluons or top quarks are non-negligible~\cite{Plehn:2008ae}.

\begin{figure}
\center
 \includegraphics[width=1.\columnwidth]{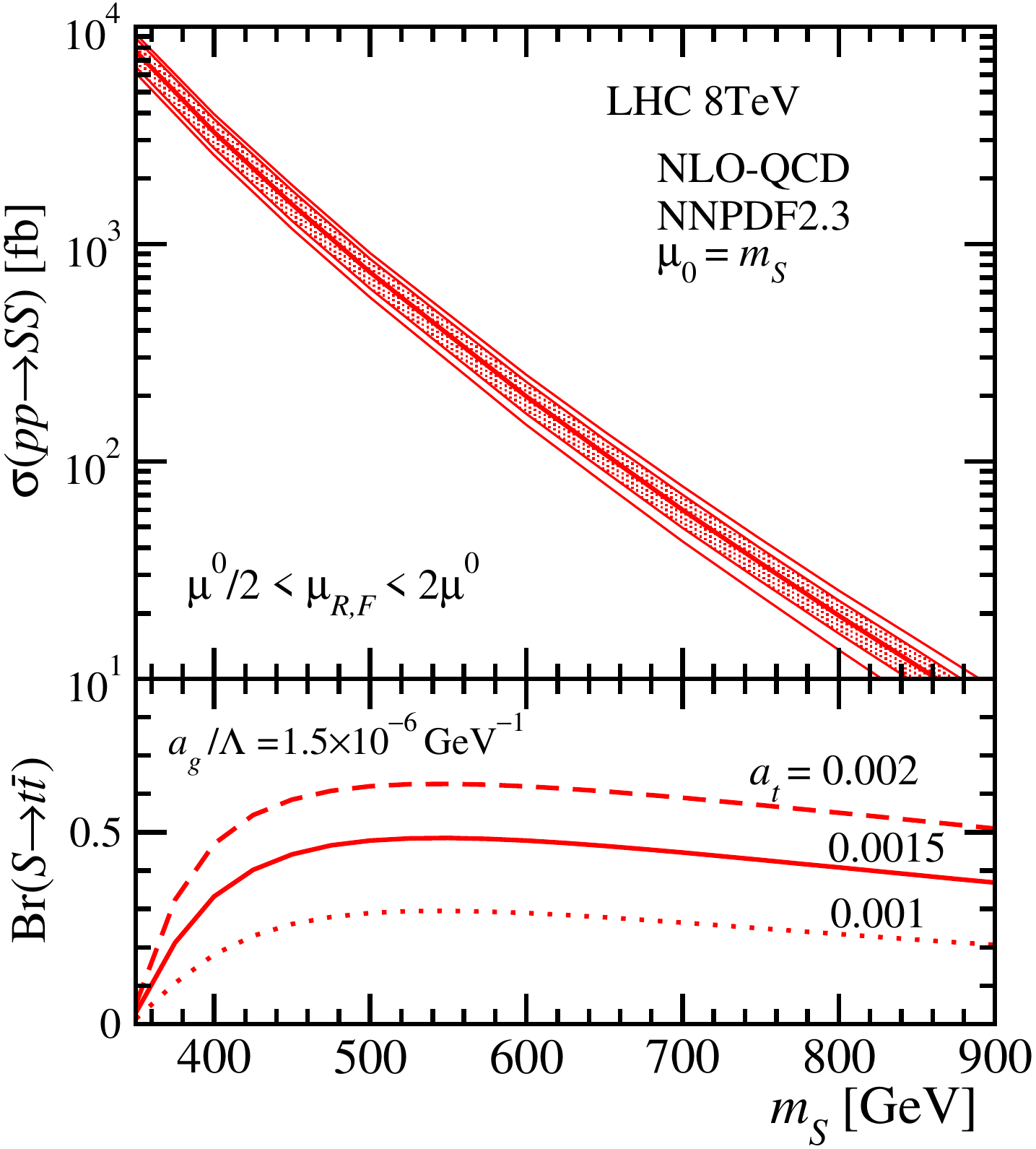}
 \caption{Total cross section for sgluon pair production at the LHC, for a
   center-of-mass energy of 8~TeV, given as a function of the sgluon mass.
   The lower panel shows the sgluon decay branching ratio
   into a top-antitop pair for different coupling strengths $a_t$.}
\label{fig:xsec}
\end{figure}

This setup corresponds to the scenarios of class II
introduced in Ref.~\cite{Calvet:2012rk}, in which the sgluon is top-philic
and only allowed to
decay into a top-antitop pair or into two gluon-induced jets. We define our
reference benchmark scenario by
\begin{align}
 &\ourat^L = \ourat^R = \ourat = 1.5\times 10^{-3}\ , \quad
 a_g = 1.5\times 10^{-3},\nonumber\\ 
 &\Lambda = 1~{\rm TeV}\ ,
\label{eq:benchmark}
\end{align}
and impose the sgluon mass $\ourmS$ to lie in the [350, 900]~GeV window.
For the phenomenological investigations performed in this work, we
additionally study deviations from this reference scenario by varying the
$\ourat$ and $a_g$ parameters in the ranges $[0.5,5]\times 10^{-3}$ and
$[1.35, 1.65] \times 10^{-3}$, respectively. 
As shown in
Ref.~\cite{Plehn:2008ae}, those numerical values can be obtained in
ultraviolet-complete supersymmetric models featuring colored superpartner masses
of about 1 or 2~TeV.

The scenarios under consideration exhibit an enhancement of the production rate,
at the LHC, of events containing four top quarks.
This is illustrated on Figure~\ref{fig:xsec} for proton-proton
collisions at a center-of-mass energy of 8~TeV. 
First, we present, on the upper
panel of the figure, the $\ourmS$-dependence of the sgluon-pair total
production cross section evaluated at the next-to-leading order (NLO) accuracy
in QCD~\cite{GoncalvesNetto:2012nt,Degrande:2014sta}. 
Second, we show, in the lower
panel of the figure, the $\ourmS$-dependence of the
sgluon branching ratios into a top-antitop system for
several values of $\ourat$.
In order to calculate the sgluon-pair total cross section at NLO
in QCD, we have followed the procedure described in
Ref.~\cite{Degrande:2014sta}. More
specifically, we have employed the {\sc FeynRules} package~\cite{Alloul:2013bka}
and its {\sc NloCT} module~\cite{Degrande:2014vpa} to generate a UFO
library~\cite{Degrande:2011ua} suitable to be used within the \amc\
framework~\cite{Alwall:2014hca}. The central curve on Figure~\ref{fig:xsec} is
then
obtained by fixing the renormalization and factorization scales to $\ourmS$ and
using the NNPDF~2.3 set of parton distributions~\cite{Ball:2012cx}, while the
uncertainty band has been derived 
by varying the two unphysical scales by a factor of two up and down with respect
to $\ourmS$ (the inner range) and 
by using all parton density replicas provided
by the NNPDF Collaboration (the outer range).

\section{Constraining top-philic sgluons from LHC Run I data}
\label{sec:lhc}
As sketched on Figure~\ref{fig:xsec}, the production of a pair of top-philic
sgluons, whose dynamics are described by the model presented in
Section~\ref{sec:model}, leads to an enhancement of LHC events containing either
two top-antitop systems, or two dijet systems, or one of each. Since in our
class of realistic scenarios, the sgluon couples to both top quarks and gluons,
the existing limits on its mass~\cite{TheATLAScollaboration:2013jha,%
Aad:2011fq,Chatrchyan:2013qha} do not directly apply and must be carefully
reinterpreted. The most stringent and model-independent constraints have
been derived in the context of an ATLAS analysis of partial LHC Run I
data~\cite{TheATLAScollaboration:2013jha}. Sgluons are in this case searched for
in same-sign dilepton events arising from the decay of a four-top ($\tttt$)
system, since such a final state has the advantage to allow for a good experimental
precision due to very low SM background. Instead of recasting this
ATLAS analysis, we focus in Section~\ref{sec:dilepton}
on its CMS counterpart~\cite{Chatrchyan:2013fea}
which benefits from the entire collision dataset at a
center-of-mass energy of 8~TeV.
Additionally, we choose to also explore the
$\tttt$ topology via its single lepton plus jets decay channel.
Although this mode exhibits considerable but well-understood background,
it allows us to exploit a larger signal branching fraction. We
reinterpret in Section~\ref{sec:singlelepton} the SM
$\tttt$ single lepton plus jets analysis of CMS~\cite{Khachatryan:2014sca}
that also covers all 8~TeV data.
Our results, that consist of the first attempt to reinterpret LHC
analyses of the full 8~TeV dataset in the aim of constraining realistic sgluon
models, are presented in Section~\ref{sec:interpretation}.

\subsection{Dilepton analysis}
\label{sec:dilepton}

The first topology chosen to be investigated in this work consists of a
same-sign dilepton signature, for which CMS has analyzed the entire recorded dataset
of collisions at a center-of-mass energy of
$\sqrt{s}=8$~TeV~\cite{Chatrchyan:2013fea}.
This CMS analysis consists of a collection of
counting experiments searching for new physics from events containing
two isolated same-sign charged leptons ($ee$, $e\mu$, $\mu\mu$) and jets.
Results are
presented in multiple search regions which are determined by requirements on the
number of selected jets ($N_{\textrm{jets}}$) with transverse momentum
$p_{\textrm{T}}>40$~GeV and pseudorapidity satisfying $|\eta| < 2.4$,
as well as on the number of jets identified as originating from the
fragmentation of a $b$-quark ($N_{\textrm{b-jets}}$), the scalar sum of the
$p_{\textrm{T}}$ of the selected jets ($H_{\textrm{T}}$) and the missing
transverse energy ($\slashed{E}_T$).
Without a knowledge of the correlations between the uncertainties on the
background expectations in the search regions, the reinterpretation of a
combination of search regions is not possible.
Therefore we utilise the results from the search region 28 (SR28) only.
This search region is chosen as its requirements can be fully emulated
using the selection efficiency parameterizations included in the
publication and closely correspond to the four-top-quark signature. 
This region is defined by requirements of
$N_{\textrm{jets}} \geq$ 4, $N_{\textrm{b-jets}} \geq$ 2,
$H_{\textrm{T}}>400$~GeV and $\slashed{E}_T>120$~GeV, imposed together with the
demand of two same-sign leptons with $p_{\textrm{T}}>$ 20~GeV and
\mbox{$|\eta|<2.4$}. Events
containing an opposite-sign same-flavor lepton pair with an invariant mass
$M_{\ell\ell}$ satisfying either $M_{\ell\ell}<12$~GeV or
\mbox{$76~\text{GeV}
< M_{\ell\ell}<106$~GeV} are also vetoed. This was implemented in the original analysis in Ref.~\cite{Chatrchyan:2013fea} to suppress background events arising from the decay of a low-mass bound state or multiboson production.

The analysis of Ref.~\cite{Chatrchyan:2013fea} contains very useful
information allowing for reinterpretation studies, following closely the
recommendations of Refs.~\cite{Kraml:2012sg,Dumont:2014tja}. 
In particular it  includes parameterizations of the selection efficiencies which allows
the signal acceptance to be estimated from generator-level information without the need
for full detector simulation.
Using these parameterizations we define an event-by-event weight
\be \label{eqn:weights}
  w_{\rm event} = \varepsilon_{H_T} \times  \varepsilon_{\slashed{E}_T} \times
    \varepsilon_{\geq\text{2b-tags}} \times \varepsilon_{\geq1 \text{SS2L}} \ ,
\ee
that uses the
individual efficiencies for the hadronic activity $H_T$ ($\varepsilon_{H_T}$),
the missing energy $\slashed{E}_T$ ($\varepsilon_{\slashed{E}_T}$),
the selection of at least two $b$-jets ($\varepsilon_{\geq\text{2b-tags}}$) and
at least one same-sign lepton pair ($\varepsilon_{\geq1 \text{SS2L}}$), this
last efficiency being obtained after considering all possible permutations among
the event leptons. In addition, a mistagging rate of a jet originating from a light
quark or a gluon as a $b$-jet of 1\% is incorporated 
in the calculation of $\varepsilon_{\geq\text{2b-tags}}$.

Using the \amc\  generator~\cite{Alwall:2014hca}, we
simulate the production, at the LHC with $\sqrt{s}=8$~TeV, of a sgluon pair that
decays into a $\tttt$ final state. Our event sample is generated inclusively in
the top decays, and we subsequently match the hard-scattering events to the
{\sc Pythia}~6 parton showering and hadronization~\cite{Sjostrand:2006za}.
Jets are reconstructed by using the anti-$k_T$ algorithm with a
radius parameter set to \mbox{$R=0.5$}~\cite{Cacciari:2008gp} as included in the
{\sc FastJet} package~\cite{Cacciari:2011ma}, while the event selection and
reweighting procedure are implemented within
the {\sc MadAnalysis}~5 framework~\cite{Conte:2012fm,Conte:2014zja}.
A signal acceptance of 0.60\% has been calculated for SM $t\bar tt\bar
t$ events using the efficiency modeling of
Eq.~\eqref{eqn:weights}.
This can be compared with the acceptance of 0.49\%
obtained by CMS for the same process~\cite{Chatrchyan:2013fea}. 
These two efficiencies agree within the 30\% uncertainty quoted on the
efficiency model.

Scanning over the parameter space introduced in Section~\ref{sec:model}, we
compute signal event yields for different $(\ourmS,\ourat)$ values by using the
relevant NLO cross section (see Figure~\ref{fig:xsec}), a luminosity of
19.6 fb$^{-1}$ and the signal acceptance derived with the efficiency model. The
obtained results are then compared to the 2 (2.21) events observed (expected) by
CMS in the
SR28 region. Considering a 30\% uncertainty on the signal yield (see above), we
use the asymptotic ${\mathrm CL_S}$ method~\cite{Cowan:2010js} as implemented in
the {\sc RooStat} package~\cite{Moneta:2010pm} to calculate a 95\% CL observed (expected) upper limit on the
number of signal events $N$ in SR28.
We have computed upper limits of 4.68 (4.89), therefore
we exclude the region of the $(\ourmS, \ourat)$ plane where $N$ is predicted to be larger than 4.68 events.

\subsection{Single lepton analysis}
\label{sec:singlelepton}
The CMS $\tttt$ single lepton analysis has been performed assuming the specific
kinematics of the SM $\tttt$ production, and it places, at the 95\%
CL an upper limit of 32~fb on the associated cross
section~\cite{Khachatryan:2014sca}. For certain values of $\ourmS$ and $\ourat$,
sgluon-induced contributions could enhance the $\tttt$ total cross
section to a value larger than 32~fb, so that constraints on the parameter space
could be extracted from the $\tttt$ single lepton analysis of CMS. However,
in order to properly estimate which regions of the parameter space are
disfavored by data, it is necessary to verify that the final state kinematics
in the decay of a sgluon pair are similar to the SM case. To this
aim, we employ a simplified matrix-element method.

Matrix-element methods are normally based upon two stages. First, the detector
level kinematics of the events of interest are translated into parton-level
kinematics via transfer functions which describe the effects of the
fragmentation, hadronization and detector reconstruction. Second, the
probabilities of predicting a specific parton-level event configuration when
assuming one
or more theoretical hypotheses are calculated on the basis of the associated
matrix elements~\cite{Kondo:1988yd,Kondo:1991dw,Kondo:1993in}. The simplified
method
used in this work relies on the fact that the event kinematics are entirely
defined by the configuration of the four top quarks, so that one could compute
the
probabilities using this information only. The decays of the top quarks and
subsequent fragmentation, hadronization and reconstruction of the final state
objects are identical for the sgluon-induced and the SM
contributions,
so that these effects are not simulated in detail. Instead, the four-momenta of
the four top quarks are smeared by sampling a Gaussian distribution with a width
corresponding to 20\% of the original values. As the events are uniquely defined
by single points in phase space, the CPU-intensive phase-space integration
normally required by matrix-element methods is avoided. Moreover, our method is
further simplified by evaluating the probabilities in the SM
hypothesis only.

In order to probe how SM-like are $\tttt$ events induced by the
decay of a sgluon pair, we make use of the \amc\ program to generate one sample
of SM $\tttt$ events and a collection of samples including sgluon
contributions for different $\ourmS$ values in the [350, 900]~GeV
range. 
We then estimate the probabilities in the SM hypothesis of the events in each sample, and compute a quantity known as the
discrimination significance ($D$)~\cite{Hocker:2007ht} that is used to assess the
similarity of the probability distributions for SM and sgluon
events. It is defined as
\be
  D = \frac{|\bar{P}_{\rm SM} - \bar{P}_{\rm Sgluon}| }
    { \sqrt{   (P^{\rm RMS}_{\rm SM})^{2}  + (P^{\rm RMS}_{\rm Sgluon})^{2} } }\ ,
\ee
where $\bar{P}_{\rm SM}$ and  $\bar{P}_{\rm Sgluon}$ are the means of the
probability distributions for the SM and sgluon cases, respectively,
and where $P^{\rm RMS}_{\rm SM}$ and $P^{\rm RMS}_{\rm Sgluon}$ are the
root-mean-squares of these probability distributions. The
numerator consists of the difference between the means of the two distributions
in question, while the denominator is the effective resolution $\sigma$ in
measuring this difference. A $D$ value of unity would hence correspond to means
differing by 1$\sigma$.

We have found that in the case of the generated sgluon samples, the
discrimination significance has a minimal value of \mbox{$D\approx 0.1$} that
is reached for $\ourmS=400$~GeV. The $D$ quantity then rises approximately linearly
to a value of \mbox{$D\approx 0.5$} for $\ourmS=900$~GeV. Furthermore, the dependence
of $D$ on $\ourat$ is found to be negligible. Consequently, the $\tttt$
kinematics for the SM and in the presence of top-philic sgluons
lighter than about 1~TeV only mildly differ, so that the parameter
space regions of our model in which the $\tttt$ cross section is
predicted larger than 32~fb are excluded.

\subsection{Interpretation}
\label{sec:interpretation}

\begin{figure}
\begin{center}
\includegraphics[width=1.\columnwidth]{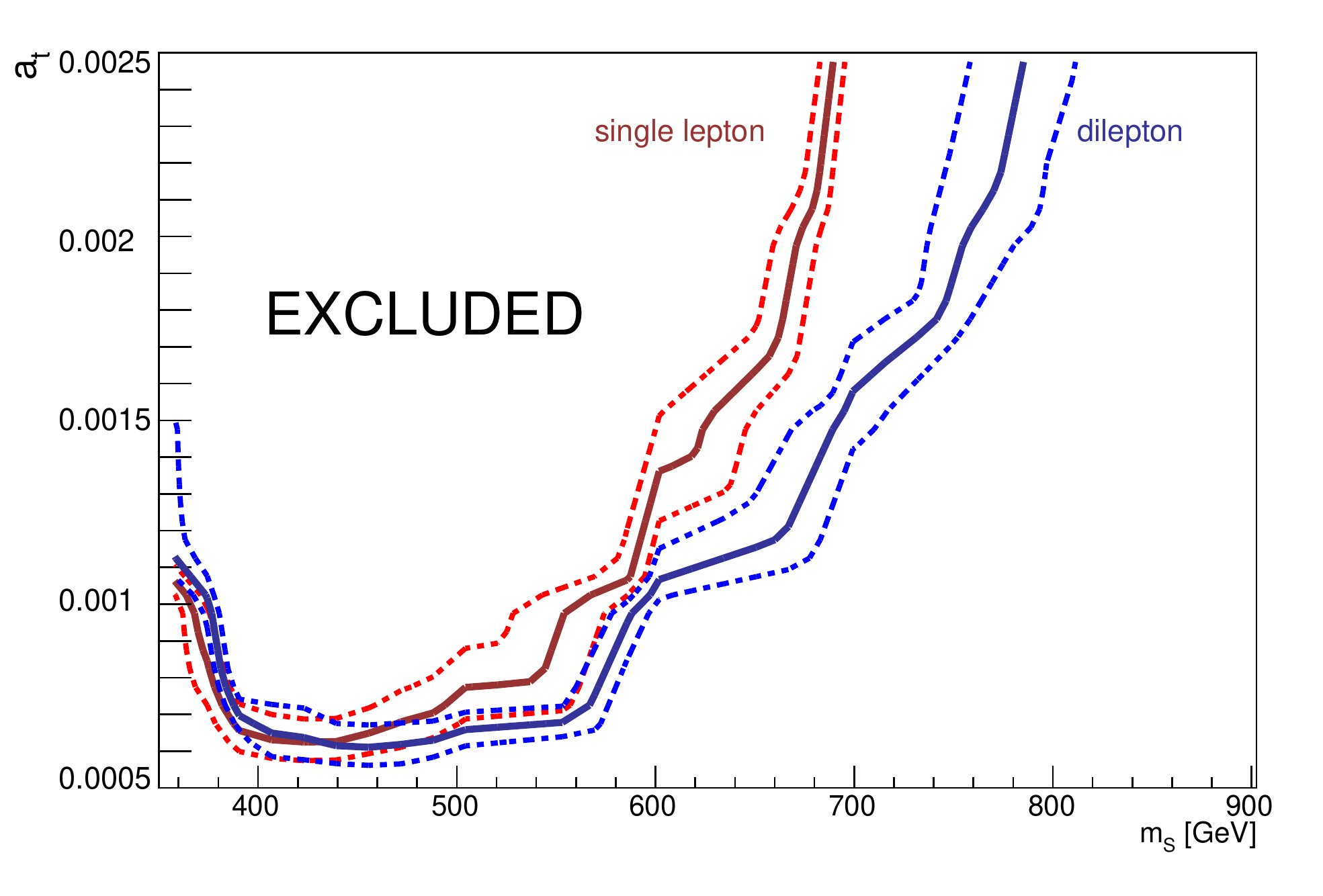}
\caption{The excluded regions in ($\ourmS, \ourat$) space derived from the single lepton analysis (red solid line) and dilepton analysis (blue solid line). In both cases, the dashed lines correspond to the exclusion regions obtained when $a_{g}$ is varied by $\pm 10\%$. }
\label{fig:exclusion}
\end{center}
\end{figure}

We present in Figure~\ref{fig:exclusion} the regions of the top-philic sgluon
parameter space that are excluded by the dilepton and single lepton analyses of
Section~\ref{sec:dilepton} and Section~\ref{sec:singlelepton}. The
results (solid lines) are represented in the $(\ourmS,\ourat)$ plane for a fixed
value of $a_g/\Lambda=1.5\times 10^{-6}$~GeV$^{-1}$ (see Eq.~\eqref{eq:benchmark}). We then
vary $a_g$ by 10\% up and down,
and show the induced variations on the constraints by dashed lines.
For a sgluon mass of about 400--550~GeV, values of $\ourat$ down to about
$0.6 \times 10^{-3}$ are excluded. This is related to the signal cross section
$\sigma(p p \to S S \to \tttt)$ which is maximal in this parameter space region.
For smaller and larger values of $\ourmS$, the sensitivity worsens
due to the decreasing sgluon branching ratio to a top-antitop system and 
the decreasing inclusive sgluon pair cross section, respectively. 

The dilepton analysis turns out
to be more constraining than the single lepton one, with constraints on $\ourmS$
ranging up to about 750~GeV for $a_t\approx 2\times 10^{-3}$. This analysis has
in addition the
advantage of being not sensitive to the signal kinematics, in contrast to the
single lepton one. We however recall that for the parameter space regions that
are reachable
with LHC collisions at $\sqrt{s}=8$~TeV, the kinematics differences between the
SM and the sgluon cases cannot be pinned down. Nevertheless, the
situation could be different for the LHC Run II expected to probe much larger
sgluon masses, as the discrimination
significance $D$ rises with $\ourmS$ (see Section~\ref{sec:singlelepton}).

\section{Conclusion}
\label{sec:conclusion}
Many new physics theories predict the existence of new colored scalar fields, named
sgluons, lying in the octet representation of QCD. These fields are
expected, at least in top-philic sgluon scenarios, to either decay to a pair of
gluons or to a top-antitop system. In this work, we have adopted
a simplified model approach describing the
dynamics of such top-philic sgluons and probed the associated
parameter space (spanned by one mass parameter $\ourmS$ and two coupling strengths
$\ourat$ and $a_g/\Lambda$) by
extracting constraints from two recent analyses of $\tttt$ events by the CMS Collaboration in
the entire LHC collision dataset at a center-of-mass energy of 8~TeV. We have found
that sgluon masses ranging up to 750~GeV are excluded for sgluon couplings to the
top quark of $\ourat=1.5\times 10^{-3}$, a value typical of non-minimal supersymmetric
scenarios with superpartner masses of about 1~TeV or 2~TeV. In the case
of smaller top-sgluon couplings of $\ourat=0.75\times 10^{-3}$ (the smallest value
which the LHC Run I is sensitive to), the sgluon mass
is imposed to satisfy $\ourmS \notin [400,550]$~GeV. 
We have finally also observed that the
picture does not drastically change with respect to $10\%$ variations of the
sgluon-gluon coupling strength whose reference value has been fixed to
$a_g/\Lambda=1.5\times 10^{-6}$~GeV$^{-1}$,
a value again typical of supersymmetric setups with superpartners around the
TeV scale.

\section*{Acknowledgments}
This work has been supported in part by the 
Belgian Federal Science Policy Office through the Interuniversity
Attraction Pole P7/37,
by the Strategic Research Program ``High Energy
Physics'' and the Research Council of the Vrije Universiteit Brussel. Further support
has been provided by the Fonds Wetenschappelijk Onderzoek through the ``Odysseus'' and ``Pegasus Marie Curie Fellowship'' programmes, by the European Commission funding channel
``Marie Curie Intra-European Fellowship for Career Development", by the Science and Technology Facilities Council in the United Kingdom and by the Th\'eorie-LHC France initiative of the CNRS/IN2P3.

\providecommand{\href}[2]{#2}\begingroup\raggedright\endgroup

\end{document}